\documentclass[11pt]{article}

\usepackage[final]{acl}

\usepackage{times}
\usepackage{latexsym}
\usepackage[T1]{fontenc}
\usepackage[utf8]{inputenc}
\usepackage{microtype}
\usepackage{inconsolata}

\usepackage{graphicx}
\graphicspath{{images/}{../images/}}

\usepackage{amsmath}
\usepackage{enumitem}
\usepackage{float}
\usepackage{tikz}
\usetikzlibrary{arrows.meta,positioning}

\usepackage{fvextra}   
\usepackage{xcolor}

\DeclareUnicodeCharacter{274C}{\texttimes} 

\fvset{frame=single, rulecolor=\color{gray!55}, framesep=4pt, breaksymbolleft={}}

\newenvironment{promptbox}[1][]
  {\par\medskip\noindent{\small\bfseries #1}\par\nobreak\vspace{2pt}\nobreak}
  {\par\medskip}

\title{Local Citation Recommendation Without Graph Signals}

\author{
  Benjamin Basseri \quad Pavlos Protopapas \quad Kemeng Zhang
  \\
  Harvard University
  \\
  \texttt{\{bbasseri,kemengzhang\}@fas.harvard.edu} \quad \texttt{pavlos@seas.harvard.edu}
  \AND
  Alberto Accomazzi \quad Stephanie Jarmak \quad Anna Kelbert
  \\
  Center for Astrophysics \textbar{} Harvard \& Smithsonian
  \\
  \texttt{\{aaccomazzi,stephanie.jarmak,anna.kelbert\}@cfa.harvard.edu}
}

\date{}

\begin{document}
\maketitle
\begin{abstract}
  Resolving correct citations can be a time-consuming task for research writers. We address local citation recommendation (LCR), where the goal is to retrieve appropriate references for a given sentence. Web tools generally depend on the existing citation graph, which may bias results toward citation count or more recent publications. LCR research using full paper texts has been limited by the lack of publicly available data. We present a new dataset, the Local Citation Recommendation Corpus (\textsc{LCRC}), containing 14{,}735 scientific statements derived from astrophysics review papers, each with one or more citation targets whose full text is accessible via API. Using this dataset we systematically compare retrieval methods and develop a citation recommendation system that avoids citation graph bias while achieving 8.0\% improvement in Recall@25 and 6.9\% improvement in NDCG@25 over dense retrieval baselines without model fine-tuning. To support reproducibility, we make our code and dataset publicly available.
\end{abstract}

\section{Introduction}
Scientific integrity has long required accurately citing prior research. Comprehensive citation, besides being expected for publication in most settings, connects research efforts across disparate teams and distant times. In recent years, scientific research in general has been subject to increased public scrutiny \citep{economist_impact_confidence_in_research_2023}. This heightens the need for thorough citations that encourage verifiability, reproducibility, and transparency in research. However, identifying and verifying appropriate citations remains a time-consuming and error-prone process, leading many writers to rely on digital tools.

Some search engines such as Google Scholar and Semantic Scholar specialize in ranking research publications, but prior work has observed that these tools favor citation counts and author metadata \citep{beel2009googlescholar, allenai_s2search_code}. These biases likely cause or exacerbate Matthew effects, where papers and authors become `popular for being popular' \citep{merton1968matthew}. The scientific community would benefit from improved and more transparent citation recommendation tools.

This work focuses on local citation recommendation (LCR), the task of predicting appropriate citations for a given sentence from a research paper. LCR research has faced challenges making results difficult to compare, reproduce, or deploy. Open datasets for this task have lacked papers' full text, likely due to copyright constraints. Lacking this data, many systems rely on metadata and the citation graph rather than a paper's true content. LCR models trained on this metadata can quickly become obsolete as new authors and papers enter the body of scientific research.

Vector stores present a logical solution to support a constantly expanding corpus and text retrieval \citep{karpukhin-etal-2020-dense-passage-retrieval}. But while a query's embedding and its citation targets may likely have high similarity, so will embeddings of different sentences citing the same source. Therefore, vector similarity alone is insufficient to discriminate between true and false positives.

LCR often resembles natural language inference (NLI), where a model attempts to predict if one text implies another. Many citing sentences $s$ cite a reference $r$ because $r$ supports the factuality of $s$, such as this one \citep{teufel-etal-2006-automatic}. But in practice, researchers cite papers for many reasons besides `$r$ implies $s$',
including to clarify, contradict, or express a unique interpretation of earlier results \citep{GarfieldEugene1996WtC}. As a result, sentences can have many different relationships to the documents they cite, relationships that are not easily resolved by vector similarity, NLI scoring, or any one single method.

We make the following contributions:
\begin{itemize}
  \item \textsc{LCRC}, a one-to-many LCR dataset derived from expert-curated review papers.
  \item A systematic evaluation of retrieval-only citation recommendation without metadata or graph signals.
  \item A mixed query representation, \textsc{Add-Prev-02+RRF} that improves Recall@25 by 8.0\% over dense retrieval, without fine-tuning.
\end{itemize}

The dataset\footnote{\url{https://huggingface.co/datasets/adsabs/LCRC}} and the code to reproduce our experiments\footnote{\url{https://github.com/benjaminb/citeline}} are both publicly available.


\section{Literature Review}

LCR systems have typically contained the following components:

\begin{itemize}
  \item Document representation
  \item Query representation
  \item Document ranking against a given query
  \item Optional filtering or reranking of top-ranked documents
\end{itemize}

with much of the research effort focused on representation strategies.

Early LCR research used various NLP techniques to construct representations. One approach computed TF-IDF vectors from passages citing a document as that document's representation \citep{heEtAlcontextawareCitationRecommendation}. Context-based Collaborative Filtering \citep{HaifengLiu2015CCFf} represented queries as binary vectors computed by citation co-occurrence with other papers in the corpus. Probabilistic approaches include \textsc{RefSeer} \citep{refseer} which adapted a translation model to produce citation likelihoods, and \textsc{Logit-Expanded} \citep{tanner-charniak-2015-hybrid} which contextualized queries with metadata including past citations and author information.

In recent years, researchers have explored neural embeddings combined with different context augmentation strategies. Early neural approaches used LSTMs to create embeddings \citep{Yang2018ALB}; subsequent work augmented embeddings with title and abstract \citep{medic-snajder-2020-improved} or modeled writer intent \citep{ma2025citation}. Transformer-based approaches include \textsc{HAtten}\citep{gu2022localcitationrecommendationhierarchicalattention}, which uses \textsc{SciBERT} \citep{beltagy2019scibert} as its embedding model, and also augments inputs with their titles and abstracts. CiteBART \citep{celik2025citebart} adapts BART \citep{lewisBARTDenoisingSequencetoSequencePretrainingNaturalLanguageGenerationTranslationComprehension2020} to generate ``(Author, Year)'' recommendations as a sequence-to-sequence task.

Another branch of research focuses on graph-based methods. Some work models citation dynamics using temporal graph neural networks \citep{holm2020longitudinal} or estimates citation impact \citep{zhang2024predicting}. \textsc{BERT-GCN} \citep{jeongcontextawarecitationrecommendationmodelBERTgraphconvolutionalnetworks2020} fine-tunes BERT \citep{devlinBERTPretrainingDeepBidirectionalTransformersLanguageUnderstanding2019} to encode queries and a graph neural network to encode the corpus. Multi-granularity preference
alignment \citep{li2025scirgcmultigranularitycitationrecommendation} combines signals derived from the citation graph with intent modeling. These approaches leverage structural and temporal information in citation networks, but are fundamentally oriented toward modeling existing citation graph dynamics, rather than identifying semantically relevant papers for a specific sentence.

A further limitation of prior LCR work lies in the datasets used for training and evaluation. Many studies rely on unreleased datasets or citation graphs derived from regular research papers, which may bias towards popularity, social networks, or other factors \citep{doughertyCitationcountsjournalimpactfactorsnotcaptureindicatorsresearchqualitybehaviouralbrainsciences2022}. Using such citation graphs as ground truth risks perpetuating these biases: when systems are evaluated against such data, it remains unclear whether they are identifying scientifically appropriate citations or simply learning to replicate existing popularity and network effects. This motivates alternative data sources, such as review papers, which reflect expert-curated citations.


\section{Task Definition \& Metrics}

\subsection{Task Definition}
In the context of research writing, we define the local citation recommendation task as retrieving the most scientifically appropriate set of documents from a corpus that should be cited for an input query.

We define the query $q$ as a sentence from a scientific manuscript, and possibly its context. Context can include additional data that helps clarify the sentence's meaning, such as preceding sentences, the paper's title, or abstract. A model performing this task aims to return the documents $D = \{d_1, \ldots, d_n\}$ which the input sentence from $q$ should cite.

For the current task we assume $D$ is non-empty and consider only nontrivial sentences, which we define as sentences that require one or more scholarly citations to support or contextualize their claims. We leave distinguishing trivial and nontrivial statements as future work.

Since the model maps queries to sets of documents, this task is a special case of document retrieval. As the task also primarily deals with natural
language, it can also be thought of as a type of semantic similarity estimation. However, it is important to note that the relationship between a research statement and the set of documents it should cite is more constrained than semantic similarity alone. If paper $c$ establishes some scientific fact which is cited by both statements $s_1$ and $s_2$ in other papers, the semantic similarity between $s_1$ and $s_2$ is likely to be high and quite possibly higher than their similarity to $c$.

Consider the case $s_1 =$ ``A superwind does not blow in low metallicity
environments'' in one paper and $s_2 =$ ``Superwinds do not blow in lower metallicity environments'' in another paper, both of which should cite $c =$ Bowen (1991) primarily for the following passage:
\begin{quote}
  ``We estimate that for $Z< 0.1Z_{\odot}$ there will usually be too little dust to produce significant dynamical effects.''
\end{quote}

Although both $s_1$ and $s_2$ should cite $c$, their near-identical wording makes it likely that $s_1$ and $s_2$ will exhibit higher cosine similarity to one another than either does to $c$. As a result, a retrieval system based purely on semantic similarity is likely to rank $s_2$ as more relevant to $s_1$ than the correct citation $c$, despite $c$ containing the supporting evidence. This illustrates why citation prediction cannot be reduced to sentence-level semantic similarity alone.

\subsection{Metrics}

A model performing the citation prediction task produces a ranked set of predicted
documents $\widehat{D} = \{\widehat{d}_1, \ldots, \widehat{d}_k\}$, often referred to as the top-$k$ documents. To evaluate the model's success, we define metrics comparing the predicted $\widehat{D}$ to the target set $D$, parameterized by $k$:

\begin{itemize}
  \item \textbf{Hitrate@$k$}: 1 if \textit{any} of the target documents
        appear in the top $k$ retrieved, 0 otherwise.
  \item \textbf{Recall@$k$}: The proportion of target documents retrieved in the top $k$.
        For example, if a query has 4 target documents
        (i.e., four reference citations) and at $k = 20$ two of those documents appear, then Recall@20 is 0.5.
  \item \textbf{NDCG@$k$}: Normalized Discounted Cumulative Gain at rank $k$ \citep{jarvelin2002cumulated}. This metric accounts for the position of relevant documents in the ranked list. Since it is common in vector-based retrieval for the same document to appear multiple times in a ranked search result due to chunking, we consider only the first occurrence of each document when computing NDCG. This is more aligned with the end goal of presenting the most likely citation candidates to the user.
\end{itemize}

Then for a dataset of $N$ queries, the mean Hitrate@$k$, Recall@$k$, and NDCG@$k$ are the sum of the individual per-query metric values divided by $N$. In Section~\ref{sec:experiments} (Experiments \& Results) we refer to these means simply as Hitrate@$k$, Recall@$k$, or NDCG@$k$ when discussing results over an entire dataset.

\section{Methodology}

To build a citation prediction model, we  constructed  a dataset of queries containing sentences requiring citations, associated metadata, and their reference citations.
We then formed a search corpus against which candidate citations are retrieved, enabling evaluation of each model's Hitrate@$k$ and Recall@$k$.
We built a labeled query dataset from review papers and a retrieval corpus from general research papers, drawn from multiple scientific literature sources.
In the experiments presented in this work, these datasets are derived from the Astrophysics Data System (ADS, \citet{Kurtz2000}).

We then segmented the corpus papers into chunks and embedded them in a vector database to support dense document retrieval. We describe our experimental setup and results in Section~\ref{sec:experiments}.

\subsection{Dataset}
To build a labeled dataset, we curated 354 digitized review papers from a scientific literature corpus spanning astrophysics, earth science, and planetary science, provided by the Astrophysics Data System. Review papers synthesize prior research and are typically written with expert-curated citation practices, making them a suitable ground truth for LCR.

From these review papers we extracted individual sentences containing inline citations and, where possible, resolved the inline citations to uniquely identifying document ID's (i.e., DOIs). We implemented an automated extraction pipeline to scale this process. First it splits the paper's body on sentence boundaries. Each sentence is then passed to an LLM agent which rejects unusable sentences such as captions, reference section entries, or sentences containing severe OCR errors. For sentences passing this filter, the agent identifies and normalizes inline citations to an array of (Author, year) tuples. The extracted (Author, year) tuples are then passed to deterministic code that maps each citation to a uniquely identifying DOI using the paper's bibliography. Sentences with any inline citations that did not fully resolve to a DOI were excluded.

For example, given the sentence:

\begin{quote}
  ``Kaiser et al (1995) (see also Luppino Kaiser 1997 , Hoekstra et al 1998 for further developments) compute the second moment within a variable aperture...''
\end{quote}

the extraction stage would yield tuples [(`Kaiser et al', 1995), (`Luppino Kaiser', 1997), (`Hoekstra et al', 1998)]. These tuples are resolved to their corresponding DOIs by searching through the review's references. The resolved papers form the target set $D$ for the query, and the inline citations are masked to form the query $q$.

Applying this pipeline produced 14{,}735 samples such that:

\begin{enumerate}[label=\alph*)]
  \item Sentences of captions, reference entries, and garbled OCR are excluded.
  \item Each sample sentence contains one or more inline citations.
  \item All inline citations are matched to their DOI.
\end{enumerate}

The dataset is provided as a JSONL file. Each line is a JSON object containing keys for the original sentence, citation target DOIs, the sentence with inline citations masked, the sentence position within the review paper, its source DOI, title, abstract, and additional fields.\footnote{\url{https://huggingface.co/datasets/adsabs/LCRC}}

\subsection{Corpus}

All dataset queries constructed cite a combined
9{,}980 distinct research papers. In our experiments we expanded the corpus to 52{,}618 distinct research papers to make the retrieval task more challenging. For the experiments presented in this work, we obtained these papers' full-text contents from ADS to form our corpus. From these records, we reconstructed each paper by concatenating the title, abstract, and body fields, intentionally omitting all other fields such as authors, affiliations, and references, in order to avoid unnecessary bias.

We implemented two forms of document representation: chunk-based and contribution-based representations. In the chunk-based representation, we used
standard text-splitting techniques to divide the papers into a corpus of 460{,}801 chunks. For the contribution-based representation, we
used a large language model to extract a paper's original findings, observations, and methodologies (see Section~\ref{sec:doc-expansion}).

\begin{figure}[htbp]
  \centering
  \begin{tikzpicture}[
      >=Stealth,
      every node/.style={font=\footnotesize},
      dsdata/.style={
          rectangle, rounded corners=2pt,
          draw=blue!50, fill=blue!8,
          minimum width=3.2cm, text width=3.0cm, align=center, inner sep=5pt},
      proc/.style={
          rectangle, rounded corners=7pt,
          draw=gray!55, fill=gray!8,
          minimum width=3.2cm, text width=3.0cm, align=center, inner sep=5pt},
      rejectd/.style={
          rectangle, rounded corners=2pt,
          draw=red!55, fill=red!8,
          minimum width=2.2cm, text width=2.0cm, align=center, inner sep=4pt},
      dsout/.style={
          rectangle, rounded corners=2pt,
          draw=blue!70, fill=blue!18,
          minimum width=3.2cm, text width=3.0cm, align=center, inner sep=5pt,
          line width=0.9pt},
    ]


    \node[dsdata] (A) at (2.5, 0) {
      \textbf{ADS Review Papers} (3,000)\\[2pt]
      \scriptsize Astrophysics $\cdot$ Earth Science $\cdot$ Planetary Science};

    \node[font=\footnotesize\bfseries, above=0.4cm of A] {Query Dataset Pipeline};

    \node[proc,   below=0.5cm of A] (B) {Sentence Splitting};

    \node[dsdata, below=0.5cm of B] (C) {
      Individual Sentences\\[2pt]
      \scriptsize\itshape ``Kaiser et al (1995) compute\\the second moment\ldots''};

    \node[proc,   below=0.5cm of C] (D) {LLM Filter Agent};

    \node[rejectd, right=0.8cm of D] (REJ) {
      \scriptsize $\times$~captions\\[2pt]
      $\times$~reference entries\\[2pt]
      $\times$~OCR errors};

    \node[dsdata, below=0.5cm of D] (E) {Filtered Sentences};

    \node[proc,   below=0.5cm of E] (F) {LLM Citation\\Extractor};

    \node[dsdata, below=0.5cm of F] (G) {
      Author--Year Tuples\\[2pt]
      \scriptsize\itshape [(`Kaiser et al', 1995),\\(`Luppino Kaiser', 1997), \ldots]};

    \node[proc,   below=0.5cm of G] (H) {DOI Resolution via Bibliography};

    \node[rejectd, right=0.8cm of H] (I) {
      \scriptsize $\times$~any citation\\[2pt]
      unresolved};

    \node[dsout,  below=0.5cm of H, minimum width=5.0cm, text width=4.8cm, xshift=0.9cm] (J) {
    \textbf{14,735 Query Samples}\\[3pt]
    \scriptsize\itshape \{query: ``[REF] compute the second moment\ldots[REF]'',\\[1pt]
    targets: [\texttt{10.1000/\ldots}, \texttt{10.2000/\ldots}]\}};

    \draw[->] (A) -- (B);
    \draw[->] (B) -- (C);
    \draw[->] (C) -- (D);
    \draw[->, red!60] (D) -- node[above, font=\scriptsize, text=red!70] {reject} (REJ);
    \draw[->] (D) -- node[left,  font=\scriptsize] {pass} (E);
    \draw[->] (E) -- (F);
    \draw[->] (F) -- (G);
    \draw[->] (G) -- (H);
    \draw[->, red!60] (H) -- node[above, font=\scriptsize, text=red!70] {reject} (I);
    \draw[->] (H) -- node[left,  font=\scriptsize] {pass} (J);

  \end{tikzpicture}
  \caption{Query dataset construction pipeline. An automated pipeline processes 3,000 ADS review papers. Sentences are split, then an LLM agent filters out captions, reference entries, and OCR errors. Inline citations are extracted as (Author, Year) tuples and resolved to DOIs via the paper's bibliography; any sentence with an unresolved citation is dropped, yielding 14,735 labeled query samples.}
  \label{fig:pipeline}
\end{figure}
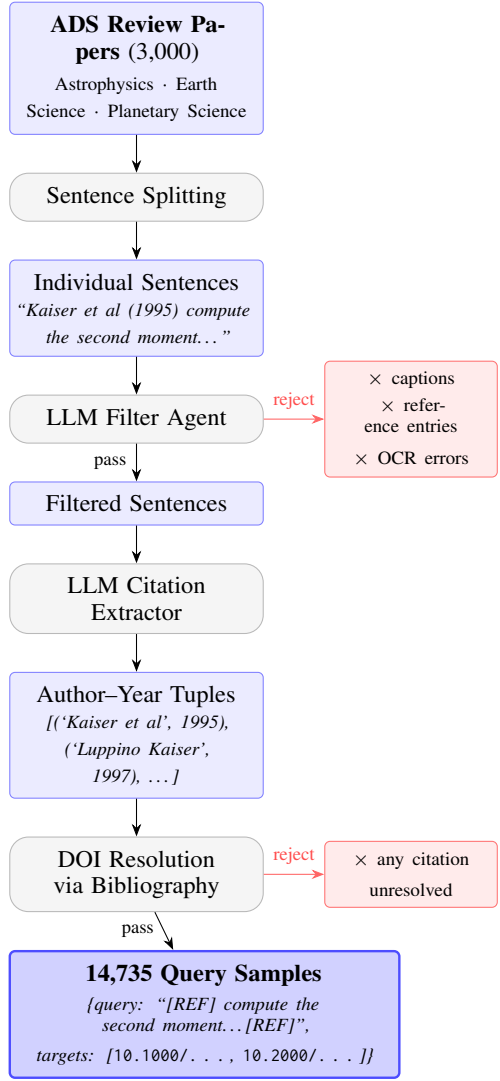


\section{Experiments and Results} \label{sec:experiments}

We organize our experiments into four distinct stages of semantic retrieval:
choice of embedder, corpus representation, query representation, and
reranking strategies. At each stage, we explored several alternatives, as
detailed in the subsections below.


\subsection{Embedding Model Selection}

To select an embedding model, we used Milvus to implement a vector database.
We chunked the research corpus and, for each embedding model evaluated,
embedded the resulting chunks.
Retrieval was then performed by computing cosine similarity between query
sentence embeddings and chunk embeddings.

Although retrieval is performed at the chunk level using cosine similarity between query and chunk embeddings, evaluation is conducted at the paper level by mapping retrieved chunks back to their source documents via DOI, and computing Hitrate@$k$ and Recall@$k$ over the resulting set of unique papers.

We evaluated \textsc{Qwen3-Embedding-0.6B}, \textsc{Qwen3-Embedding-8B} \citep{zhang2025qwen3embeddingadvancingtext}, and \textsc{BGE-large-en-v1.5} \citep{bge} models for their high-quality, general-purpose embeddings. Additionally, we evaluated several LLMs fine-tuned on
astrophysics data as potential embedding models: \textsc{AstroBERT} \citep{grezesBuildingastroBERTlanguagemodelAstronomyAstrophysics2021a}, \textsc{AstroLlama} \citep{nguyenAstroLLaMASpecializedFoundationModelsAstronomy2023}, \textsc{AstroSage} \citep{astrosage}, \textsc{NASA-IBM-st.38m} \citep{bhattacharjeeINDUSEffectiveEfficientLanguageModelsScientificApplications2024}, and \textsc{Specter2} \citep{singh2023scirepevalmultiformatbenchmarkscientific}.

We find that \textsc{Qwen3-Embedding-8B} achieves the best hitrate and recall for all values of $k$, with \textsc{Qwen3-Embedding-0.6B} ranking second-best despite being a significantly smaller model. These general-purpose embedding models, along with \textsc{bge-large-en-v1.5}, outperform the astrophysics fine-tuned models overall (see Table~\ref{tab:embedders}).



\begin{table*}[hbtp]
  \centering
  \small
  \setlength{\tabcolsep}{4pt}
  \begin{tabular}{l|rrr|rrr|rrr}
    \hline
                                         & \multicolumn{3}{c|}{\textbf{Hitrate}} & \multicolumn{3}{c|}{\textbf{Recall}} & \multicolumn{3}{c}{\textbf{NDCG}}                                                                                                           \\
    \textbf{Embedder}                    & @10                                   & @25                                  & @50                               & @10             & @25             & @50             & @10             & @25             & @50             \\
    \hline
    \textsc{astroBERT}                   & 0.1002                                & 0.1621                               & 0.2256                            & 0.0829          & 0.1372          & 0.1946          & 0.0472          & 0.0613          & 0.0731          \\
    \textsc{astrollama}                  & 0.3221                                & 0.4347                               & 0.5292                            & 0.2844          & 0.3938          & 0.4847          & 0.1881          & 0.2163          & 0.2349          \\
    \textsc{AstroSage-8B}                & 0.1898                                & 0.2611                               & 0.3237                            & 0.1622          & 0.2282          & 0.2873          & 0.1068          & 0.1240          & 0.1361          \\
    \textsc{bge-large-en-v1.5}           & 0.5623                                & 0.6617                               & 0.7255                            & 0.5126          & 0.6160          & 0.6844          & 0.3868          & 0.4139          & 0.4281          \\
    \textsc{nasa-ibm-st.38m}             & 0.5241                                & 0.6269                               & 0.7020                            & 0.4761          & 0.5796          & 0.6585          & 0.3560          & 0.3832          & 0.3996          \\
    \textbf{\textsc{Qwen3-Embedding-8B}} & \textbf{0.6640}                       & \textbf{0.7570}                      & \textbf{0.8126}                   & \textbf{0.6098} & \textbf{0.7119} & \textbf{0.7757} & \textbf{0.4801} & \textbf{0.5072} & \textbf{0.5206} \\
    \textsc{specter2}                    & 0.3634                                & 0.4706                               & 0.5513                            & 0.3249          & 0.4278          & 0.5070          & 0.2302          & 0.2569          & 0.2731          \\
    \hline
  \end{tabular}
  \caption{\textbf{Embedding Model Comparison.} Hitrate, Recall and NDCG at ranks 10, 25, and 50 for evaluated embedders. General-purpose embedding models \textsc{Qwen3-Embedding-8B} and \textsc{bge-large-en-v1.5} consistently outperform astrophysics fine-tuned LLMs.}
  \label{tab:embedders}
\end{table*}

\subsection{Corpus Representation} \label{sec:doc-expansion}

We attempted to find a more effective representation
for the reference corpus by tasking an LLM to write
an array of sentences comprising each research
paper's findings, scientific contributions, and
methodologies, similar to the document expansion
technique described by \citet{nogueira2019document}.
This sentence array aimed to provide a more
concise summary of the paper's content that is
most likely to be cited. We collectively refer to
all such citable statements as ``contributions''.
We used DeepSeek-v3.2 \citep{deepseekai2025deepseekv32pushingfrontieropen} for its advanced reasoning capabilities to produce the contributions-based document expansion, constructing a new corpus using a simple prompt (v.~1) and a more advanced prompt (v.~2). The
simple prompt resulted in a corpus of 89{,}860
statements, while the advanced prompt resulted
in a corpus of 299{,}286 statements. However, the
baseline chunk-based representation of the corpus outperformed both contribution-based variants. Results are shown in Fig~\ref{fig:doc-rep}, and the extraction prompts are included in Appendix~\ref{app:prompts}. One possible explanation is that citation decisions are often driven by narrow, localized evidence—such as specific definitions, caveats, or technical claims—rather than high-level summaries. Chunking may better preserve these fine-grained cues, whereas contribution-level summaries could abstract away details that are important for aligning a citing sentence with its source.\footnote{
  We also experimented with summarizing individual chunks using an LLM and appending these summaries prior to embedding. This approach did not improve retrieval performance in preliminary experiments and was not pursued further.
}

\begin{figure}[t]
  \centering
  \includegraphics[width=0.75\columnwidth]{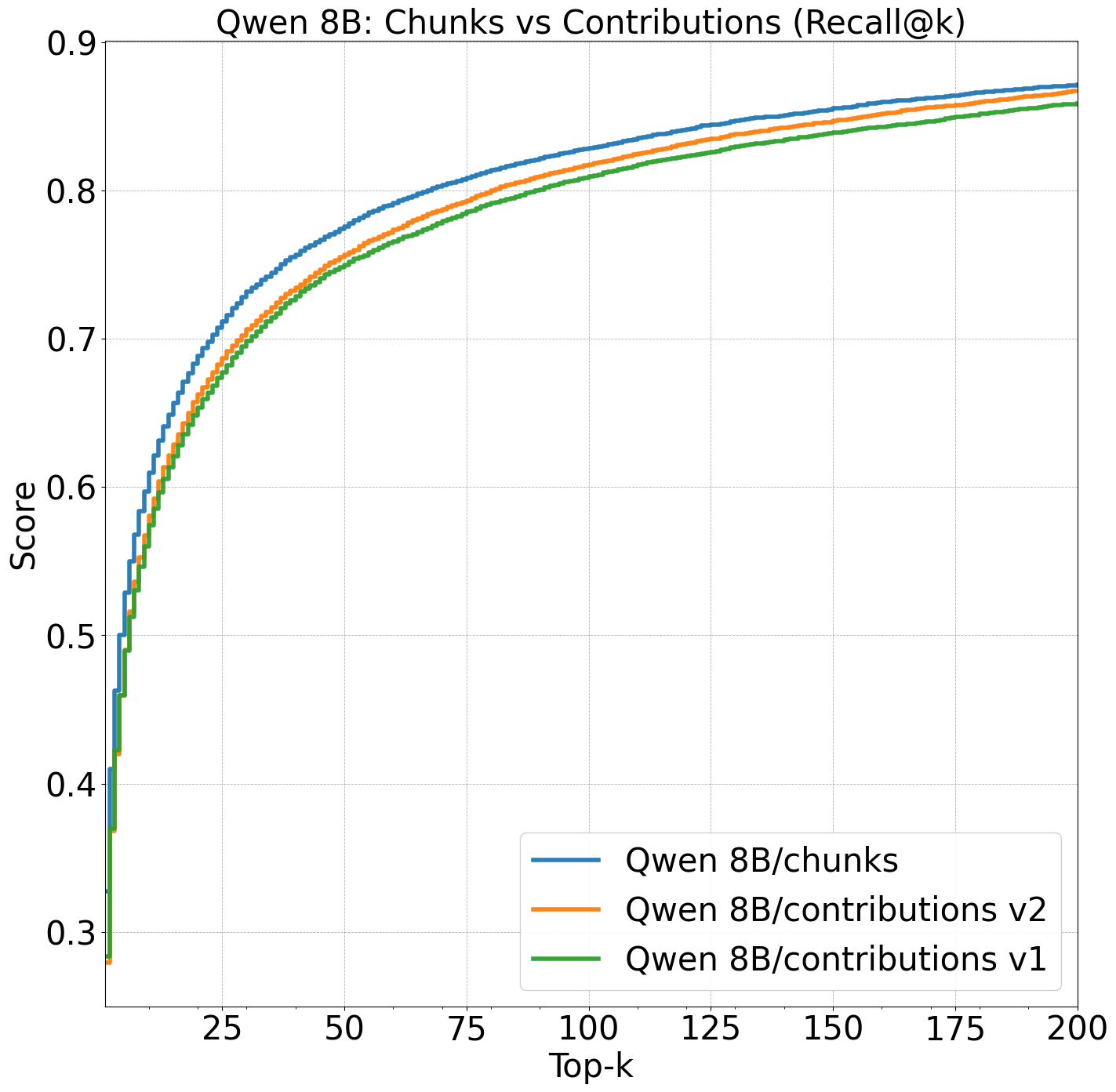}
  \caption{Comparison of document representation strategies. Baseline chunking outperformed LLM-extracted contributions (v1: simple prompt; v2: engineered prompt). Hitrate@$k$ and NDCG@$k$ showed similar trends.}
  \label{fig:doc-rep}
\end{figure}

\subsection{Query Representation}

To better contextualize the input query, we attempted several types of query context management before embedding. These include prepending the title or abstract from the query's original paper, or the previous $n$ sentences for $n \in \{1, 2, 3, 5, 7\}$. We initially found that representing the input
query sentence by prepending its previous three sentences, \textsc{Add-Prev-3}, was the best-performing strategy at many values of $k$, suggesting that this additional context helped the embedder better encode the underlying meaning of the query sentence.

After some informal error analysis of the \textsc{Add-Prev-3} strategy, we found that some queries benefited from the additional context but
others did not. This led to the development of a `mixed query representation' strategy.

In mixed query representation, we represent the input query with multiple distinct representations. Specifically, we take the original sentence and the sentence appended to its previous two sentences, then embed them in parallel. Each vector is used to perform cosine similarity search against the database, retrieving the top $k // 2$ records. In other words, to retrieve a total of $k$ results, we retrieve half from the query's original text and the other half from the augmented query. We merge the two resulting sets by interleaving by rank: for chunks $(c_1, \ldots, c_{k//2})$ returned by the unmodified query and $(e_1, \ldots, e_{k//2})$ from the augmented query, the final ranking is $(c_1, e_1, c_2, e_2, \ldots, c_{k//2}, e_{k//2})$. This interleaving outperformed sorting by cosine similarity alone.

We call this query representation strategy \textsc{Add-Prev-02} since it combines the query sentence as-is (i.e., with zero previous sentences) with the query sentence plus its previous two sentences. This strategy significantly outperformed the baseline \textsc{Identity} and all other query representations at all values of $k$.

One possible explanation is that, while additional context can help disambiguate certain citation intents, it may also introduce topic drift. In review papers, preceding sentences often summarize broader background or adjacent results that are only loosely related to the specific claim being cited. As a result, uniformly expanding all queries with surrounding context may dilute the semantic signal most relevant for citation selection.

The mixed query representation may mitigate this trade-off by retaining both a high-precision signal (the isolated sentence) and a higher-recall signal (the context-augmented version). By combining these complementary views through rank interleaving or fusion, the retrieval process can preferentially surface candidates that are robust across representations, while reducing the impact of misleading context.

\begin{table*}[hbtp]
  \centering
  \small
  \setlength{\tabcolsep}{4pt}
  \begin{tabular}{l|rrr|rrr|rrr}
    \hline
                                      & \multicolumn{3}{c|}{\textbf{Hitrate}} & \multicolumn{3}{c|}{\textbf{Recall}} & \multicolumn{3}{c}{\textbf{NDCG}}                                                                                                           \\
    \textbf{Method}                   & @10                                   & @25                                  & @50                               & @10             & @25             & @50             & @10             & @25             & @50             \\
    \hline
    \textsc{Identity}                 & 0.6640                                & 0.7570                               & 0.8126                            & 0.6098          & 0.7119          & 0.7757          & 0.4801          & 0.5072          & 0.5206          \\
    \textsc{Add-Title}                & 0.6546                                & 0.7562                               & 0.8225                            & 0.6012          & 0.7107          & 0.7862          & 0.4647          & 0.4936          & 0.5094          \\
    \textsc{Add-Abstract}             & 0.4422                                & 0.5798                               & 0.6807                            & 0.3991          & 0.5350          & 0.6387          & 0.2725          & 0.3077          & 0.3291          \\
    \textsc{Add-Title+Prev-3}         & 0.6085                                & 0.7268                               & 0.8012                            & 0.5578          & 0.6812          & 0.7620          & 0.4091          & 0.4415          & 0.4584          \\
    \textsc{Add-Prev-3}               & 0.5944                                & 0.7169                               & 0.7953                            & 0.5442          & 0.6707          & 0.7554          & 0.3984          & 0.4317          & 0.4493          \\
    \textsc{Add-Prev-5}               & 0.5389                                & 0.6692                               & 0.7561                            & 0.5928          & 0.6238          & 0.7153          & 0.3527          & 0.3870          & 0.4059          \\
    \textsc{Add-Prev-7}               & 0.4973                                & 0.6289                               & 0.7225                            & 0.4539          & 0.5847          & 0.6823          & 0.3175          & 0.3518          & 0.3719          \\
    \textsc{Add-Prev-02}              & 0.7157                                & 0.8094                               & 0.8681                            & 0.6580          & 0.7637          & 0.8329          & 0.5069          & 0.5351          & 0.5495          \\
    \textbf{\textsc{Add-Prev-02+RRF}} & \textbf{0.7190}                       & \textbf{0.8140}                      & \textbf{0.8735}                   & \textbf{0.6624} & \textbf{0.7692} & \textbf{0.8383} & \textbf{0.5135} & \textbf{0.5420} & \textbf{0.5565} \\
    \hline
  \end{tabular}
  \caption{\textbf{Query Expansion Comparison.} Hitrate, recall, and NDCG using \textsc{Qwen3-Embedding-8B} and various query expansion methods.}
  \label{tab:query-expansions}
\end{table*}

\subsection{Reranking by Reciprocal Rank Fusion}

Reciprocal rank fusion (RRF) reranks the top-$k$ search results by forming a weighted sum of their \textit{reciprocal} ranks (so that lower-ranked documents receive smaller contributions) with respect to multiple rerankers $R_1, \ldots, R_n$, using weights $w_1, \ldots, w_n$, plus a smoothing parameter $m$ \citep{cormack2009reciprocal}.

$$
  \text{RRF}(q, d)
  =
  \frac{w_1}{m + R_1(q, d)}
  + \ldots +
  \frac{w_n}{m + R_n(q, d)}
$$

We tested two rerankers, BM25 \citep{robertson2009probabilistic} and
\textsc{num\_retrievals}. BM25, a well-known
text similarity function, uses term frequency and
inverse document frequency to compute a
similarity score between a query and a document
\citep{robertson2009probabilistic}. For
\textsc{num\_retrievals}, we hypothesized that a
larger number of times the same source document
appeared in the top-$k$ chunks retrieved by vector
similarity search may indicate stronger
query--document similarity. To test this
hypothesis, we defined the \textsc{num\_retrievals}
function as the number of times a source paper
appears in the search results, as identified by its
DOI.

After performing a grid search over these rerankers with various weights, we find the best performing RRF function uses BM25 with a weight of 0.4, fused with the original search results rank weighted at 0.6:

\begin{equation*}
  \begin{split}
    \text{RRF}^*(q, d) = {} & \frac{0.6}{m + R_\text{cos}(q, d)} \\
    & + \frac{0.4}{m + R_{\text{BM25}}(q, d)}
  \end{split}
\end{equation*}
where $R_{\cos}(q, d)$ is the document chunk's original rank with respect to the query (by cosine similarity) and $R_{\text{BM25}}(q, d)$ is the document chunk's rank when reranking by BM25 (with parameters $k_1 = 1.5, b = 0.75$), and $m = 60$. Meanwhile, rank fusion with \textsc{num\_retrievals} did not improve scores.


\subsection{Evaluating Citation Count or Publication Date Bias} We evaluated our dataset using rerankers based on citation count and publication age to assess whether there appeared to be a preference for higher citation counts or more recent publications in our data. To do so, we defined two rerankers:
\begin{align*}
  \textsc{recency}        & := -\log (y + 1) \\
  \textsc{log\_citations} & := \log(c + 1)   \\
\end{align*}
where $y$ is computed as length in years between query's publication date and the chunk's publication date, and $c$ is the citation count of the chunk's source paper. We add 1 to each logarithm to avoid evaluating $\log(0)$.

We conducted a grid search by combining each reranker with the original ranks from
\textsc{Add-Prev-02}/\textsc{Qwen3-Embedding-0.6B} results. We searched over non-negative weight combinations in increments of $0.1$. At all weight combinations evaluated, RRF with either of these metrics generally caused hitrate and recall to decrease for most values of $k$, suggesting that the review papers did not
exhibit a strong preference for higher citation counts or more recent publication among semantically similar papers.

\begin{figure*}[htbp]
  \centering
  \includegraphics[width=0.48\textwidth]{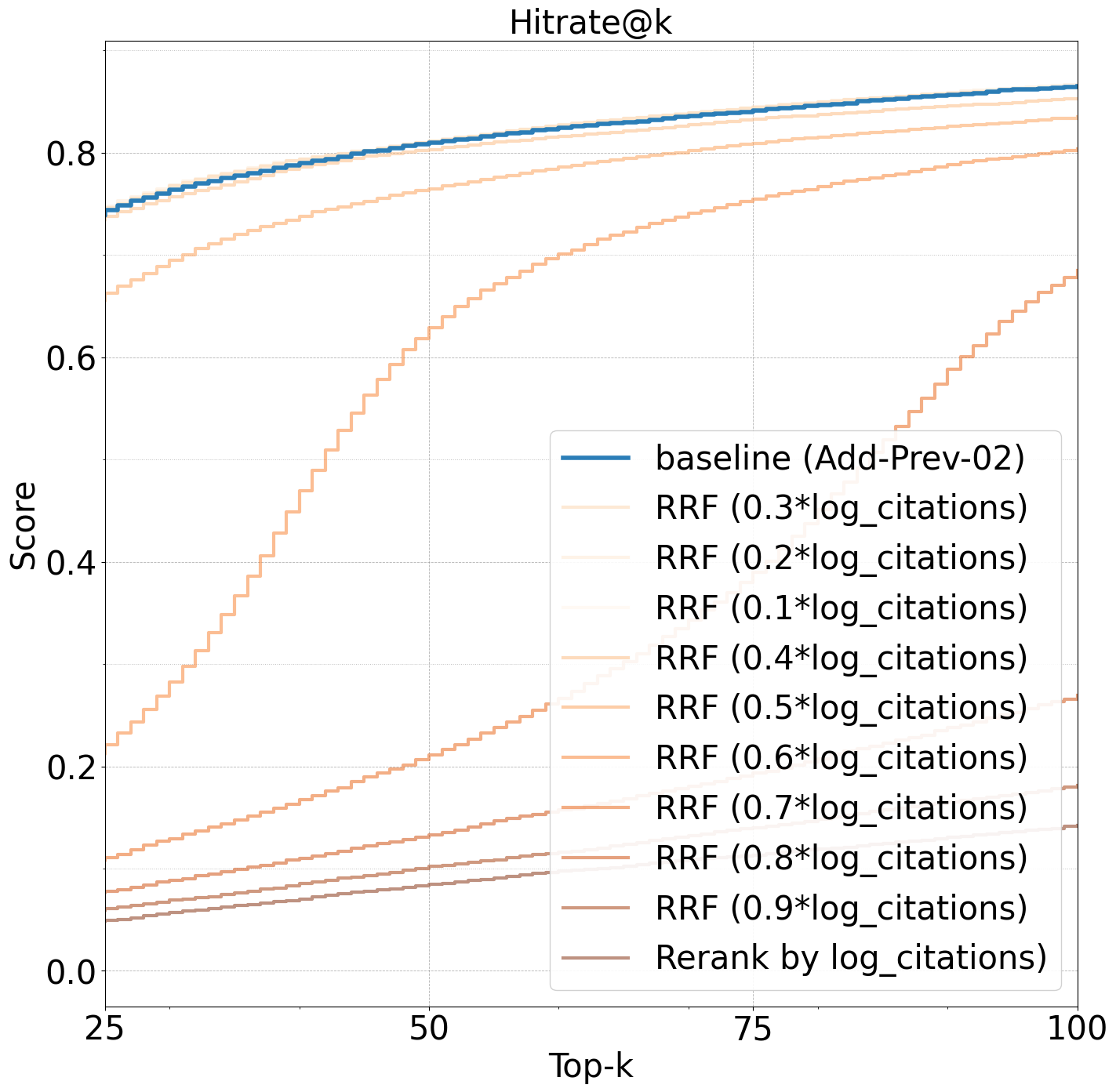}
  \hfill
  \includegraphics[width=0.48\textwidth]{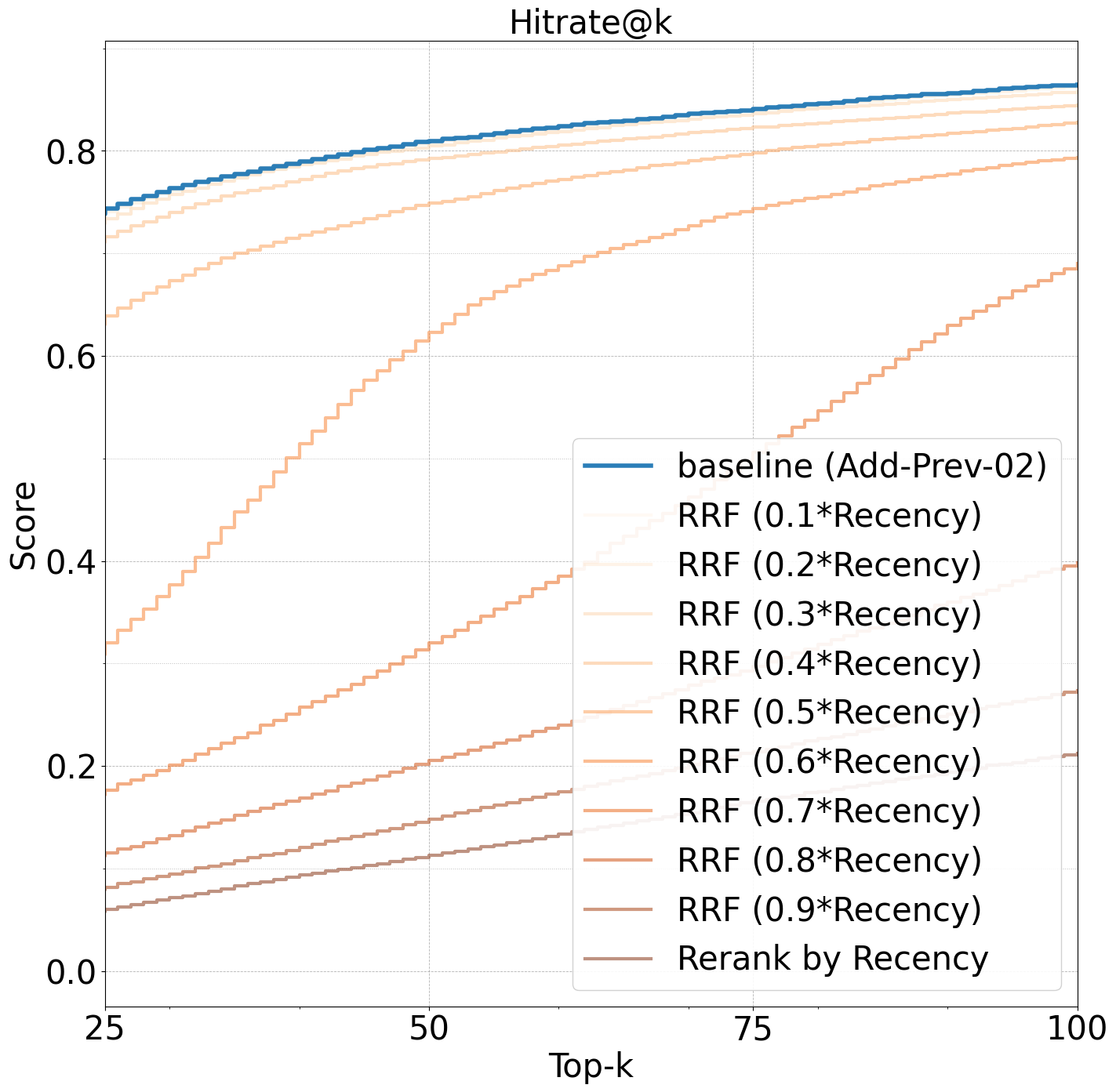}
  \caption{Grid search results for reciprocal rank fusion on original ranks from \textsc{Add-Prev-02} with \textsc{Log\_Citations} and \textsc{Recency}. The top-scoring weights at best matched the original ranking from \textsc{Add-Prev-02} for most values of $k$. Applying higher weights to \textsc{Log\_Citations} and \textsc{Recency} tended to reduce hitrate and recall.}
  \label{fig:rrf_bias}
\end{figure*}

One possible explanation is that review articles are intentionally written to cite foundational and semantically appropriate work, rather than prioritizing the most recent or most highly cited publications. As a result, rerankers based on citation count or publication recency may introduce signals that are weakly aligned—or even misaligned—with the underlying citation intent, leading to degraded retrieval performance (see Fig~\ref{fig:rrf_bias}).

\section{Conclusions \& Future Work}

In this paper, we release a new dataset for multi-label citation prediction with comprehensive metadata for retrieving citation contexts. We hope this encourages more work that focuses on the content of scientific statements rather than authorship or citation counts.
We also introduce an effective context management technique, \textsc{Add-Prev-02}, which performs citation prediction significantly better than baseline vector similarity. This is consistent with classic IR research showing multiple query representatives can improve retrieval \citep{belkineffectmultiplequeryrepresentationsinformationretrievalsystemperformance1993, belkinCombiningevidencemultiplequeryrepresentationsinformationretrieval1995}. Our strategy can be viewed as an instance of this principle in a dense retrieval setting, using natural context rather than generated query variants.

Furthermore, we identify \textsc{Qwen3} as a
top-performing embedding model for the citation
prediction task, and we identify a query
representation and rank fusion strategy that collectively improve
Hitrate@25 by 7.5\% and Recall@25 by 8.0\%
above the baseline approach. Our experiments
suggest that LLMs specifically trained for embedding tended to outperform LLMs fine-tuned on astrophysics text. To represent documents in the corpus, simple chunking outperformed the summarization methods attempted. We also show that expanding the query with additional context produced limited results when used as the only
input. However, when used in parallel with the original query, merged results
significantly improved. Finally, our reranking experiments found that rank fusion between BM25 and the raw semantic similarity ranking achieved better results than reranking by BM25 or other methods.

We find that \textsc{Qwen3} embedding models outperformed all other models fine-tuned on astrophysics data. This may be due to \textsc{Qwen3}'s contrastive loss function, which enables it to learn high-quality document embeddings. Fine-tuning astrophysics-specific models with a contrastive loss, or further fine-tuning \textsc{Qwen3} on domain-specific data, may improve performance by combining the strengths of both approaches.

The majority of our model's improvement comes from the mixed query representation strategy. Future work may explore improved combinations of query representations and evaluate whether these techniques generalize to other scientific domains.

\section*{Limitations}

Our dataset excludes queries with significant amounts of math notation since the digitized text often fails to accurately represent the original expressions. Short of re-scanning the corpus with more powerful OCR, additional work is required to obtain accurate representations or address inaccurately digitized text.

In this paper, we focused on ``nontrivial'' inputs: sentences that require at least one citation. Not all statements require a formal citation, so additional work is needed to distinguish between such statements.

\section*{Acknowledgments}

The computations in this paper were run on the FASRC Cannon cluster supported by the FAS Division of Science Research Computing Group at Harvard University. This research has made use of NASA's Astrophysics Data System Bibliographic Services.

\bibliography{custom_bib_revised}
\appendix
\onecolumn
\raggedbottom
\section{Agent \& Prompts}
\label{app:agent}

To construct our dataset we built an agent to implement the following pipeline:
\begin{enumerate}
  \item Given an input sentence, identify if it is `valid' (e.g. no serious transcription errors, not a caption)
  \item If the sentence is valid, identify all substrings that constitute inline citations
  \item Write out a version of the input sentence with those substrings masked
  \item Write out an array of (author, year) tuples for each inline citation identified
  \item Execute deterministic code that maps each (author, year) tuple to the unique DOI that represents the cited paper
\end{enumerate}

The prompts used in the agent are presented below:

\begin{promptbox}[Sentence Validity Prompt]
  \begin{Verbatim}[breaklines=true, fontsize=\small]
    You are an expert data analyst in astrophysics and earth sciences. You are building a dataset of snippets from research papers. Usable snippets will contain complete factual or scientific statements and possibly an inline citation. But snippets from a research paper's reference section or captions with little semantic content will NOT be usable.

    Your task is to read the snippet below, write a reason for if the snippet appears usable, then write True if it is usable or False if not usable.

    ## Rules
    - Write false for snippets that are primarily captions from a figure, table, or image, with no explanatory text.
    - Write false for snippets that look like they have a lot of OCR errors or math gibberish
    - Write true otherwise, REGARDLESS OF TYPOS or phrasing

    ## Examples
    Input: "Fig. 12 shows the K -band number counts from an accumulated exposure of 1.5 h (6 h of data) on a single tile in the ELAIS-N1 field."
    Output: {{"reasoning": "Although this looks like a caption to a figure, it is a complete factual statement", "label": true}}

    Input: "In relatively uncrowded regions, the aperture photometry reaches close to the target depths at J , H and K (see results in Warren et al. 2007)."
    Output: {{"reasoning": "This refers to photometry metrics in specific regions and is a complete factual statement", "label": true}}

    Input: "Casali et al. (2007) describe the survey instrument, WFCAM."
    Output: {{"reasoning": "Although this is a short sentence it states that Casali et al. (2007) will contain a description of the WFCAM instrument. So it's a factual statement.", "label": true}}

    Input: "Figure 2 Gas temperature ( red circles ); cumulative gas mass ( blue squares )"
    Output: {{"reasoning": "This appears to be a figure caption that just labels the parts of the figure. Not a scientific or usable statement.", "label": false}}

    Input: "Table 1 Observational evidence for AGN feedback"
    Output: {{"reasoning": "This looks like just a title for a table, not a complete scientific statement", "label": false}}

    Input: "Allen C. W. , 1973 , Astrophysical Quantities , 3rd edn . Univ. of London , Athlone Baraffe I."
    Output: {{"reasoning": "This looks like just a citation, possibly from a reference section. Not a complete factual or scientific statement.", "label": false}}

    Input: "It also hosts a Compton-thick AGN in the Western component, observed directly in hard X-rays (Della Ceca et al. 2002 ; Ballo et al. 2004 )."
    Output: {{"reasoning": "Describes a scientific observation", "label": true}}

    Based on the input sentence, provide your reasoning and a boolean 'isValid' flag.

      {format_instructions}

    Input:
    {sentence}
  \end{Verbatim}
\end{promptbox}

\begin{promptbox}[Inline Citation Extraction Prompt]
  \begin{Verbatim}[breaklines=true, fontsize=\small]
    System: You are a strict extractor of inline citations from snippets. You NEVER normalize, format, or clean up the text. You only extract exact spans from the snippet, with no added or missing characters, punctuation, or whitespace.

    You are an expert data analyst in astrophysics and earth sciences. You are building a dataset of snippets from research papers. In each snippet you must identify any substrings that are INLINE CITATIONS.

    Your task is to read the snippet below, think step by step about what substrings (if any) are INLINE CITATIONS, then write out those substrings as an array. If the snippet has no citations, write an empty array.

    ## Definition of INLINE CITATION:
    An inline citation is any span of text in the snippet that refers to other authors' work, including publication year or pre-publication status (e.g. 'submitted')

    ## INLINE CITATION formats:
    Inline citations can appear in various formats
    - Any sequence that looks like a capitalized name (optionally followed by 'et al.' or 'and Coauthor') immediately followed by a four-digit year (with optional letter suffixes) inside parentheses
    - Parenthetical: (Smith 1994), (Smith and Jones 1994), (Smith et al. 1994)
    - Non-parenethetical: Smith 1994, e.g. Smith 1994, Smith and Jones 1994
    - Only year in parentheses: Smith (1994), e.g. Smith (1994 ), Smith and Jones ( 1994 )
    - Multiple: (Smith 1994; Jones 1995, Brown 2000a,b)
    - With "submitted" or "in prep": (Kopp et al. submitted), (Doe 2020 in preparation)
    - Narrative: Smith (1994), Smith and Jones (1994), Smith et al. (1994)
    - Loose year mentions in parentheses: (1994), (2000a), (2000a,b)
    - Inconsistent whitespace: ( Smith 1994), ( Smith 1994 ), Smith (1994 )

    ## Not citations:
    - Expressions not including a year: ( STS 06), ( Fig. 11), (Marine Isotopic Stage, MIS 5e at ~ 122 ka)

    ## IMPORTANT:
    In your substrings
    - DO NOT alter the spacing, commas, or punctuation of any citation. You must PRESERVE THE ORIGINAL WHITESPACE, PUNCTUATION, AND FORMATTING EXACTLY as it appears in the snippet.
    - DO NOT add any parentheses not in the original snippet
    - DO NOT remove any parentheses that are in the original snippet
    - If the citation does not have parentheses in the original snippet, you must not add them in the output — even if it looks like other citations that use them.
    - DO NOT "clean up" spacing or insert/remove punctuation — your substring must be a **verbatim substring** of the original snippet.
    - You are **NOT** formatting citations — you are extracting them **exactly as written**.
    - If the citation is ( Smith 1994 ), your output must be ["( Smith 1994 )"], **not** ["(Smith 1994)"].


    ## Examples:
    Snippet: "These are minor species, with abundances of 1.5 10 12 , 1.7 10 9 , and 7 10 8 kg N, respectively ( Ussiri and Lal, 2013 )."
    Substring: ["( Ussiri and Lal, 2013 )"]
    ❌ WRONG: ["(Ussiri and Lal, 2013)"]

    Snippet: "Focusing on 353-GHz data at high Galactic latitudes, the E and B angular power spectra of dust polarization were constrained in the multipole range of 40l600 ( Adam et al. 2016 )."
    Substring: ["( Adam et al. 2016 )"]
    ❌ WRONG: ["(Adam et al., 2016)"]

    Snippet: "Proton blazar models ( Mannheim,1993 ) were initially motivated by the search of sources of ultra-high-energy cosmic rays."
    Substring: ["( Mannheim,1993 )"]
    ❌ WRONG: ["(Mannheim, 1993)"]

    Snippet: "Methods for mixing below the convection zone are well understood ( Brun, Turck-Chièze Zahn 1999 , Charbonnel Talon 2005 )."
    Substring: ["( Brun, Turck-Chièze Zahn 1999 , Charbonnel Talon 2005 )"]
    ❌ WRONG: ["Brun, Turck-Chièze Zahn 1999", "Charbonnel Talon 2005"]

    Snippet: "Like Caffau et al. (2008a) , we have similar findings."
    Substring: ["Caffau et al. (2008a)"]

    Snippet: "Momentum balance gives an expression ( Fabian 1999 ; Di Matteo, Wilman Crawford 2002 ; King 2003 , 2005 )"
    Substring: ["( Fabian 1999 ; Di Matteo, Wilman Crawford 2002 ; King 2003 , 2005 )"]
    ❌ WRONG: ["(Fabian 1999)", "(Di Matteo, Wilman Crawford 2002)", "(King 2003 , 2005)"]

    Snippet: "This first contribution describes the cluster dark matter halo (and implicitly assumes that the ICM follows a similar distribution; see Semboloni et al. ( 2011 ) for a study where this assumption is relaxed)."
    Substring: ["Semboloni et al. ( 2011 )"]
    ❌ WRONG: ["( Semboloni et al. ( 2011 ) )"]

    Snippet: "Sweatman (2021) ignored data from carbon spherules generated by YDIH proponents."
    Substring: ["Sweatman (2021)"]
    ❌ WRONG: ["( Sweatman 2021 )"]

    Snippet: "In the early Universe, when the metal content was extremely low, enrichment by a single supernova could dominate preexisting metal contents (e.g., Audouse Silk 1995 ; Ryan, Norris Beers 1996 )."
    Substring: ["Audouse Silk 1995 ; Ryan, Norris Beers 1996"]

    Snippet: "Caffau and Brun (2009) made several imaging advances primarily to identify quasars"
    Substring: ["Caffau and Brun (2009)"]
    ❌ WRONG: ["( Caffau and Brun 2009 )"]

    Snippet: "Turck-Chieze(2005) identified when metal content was extremely low"
    Substring: ["Turck-Chieze(2005)"]
    ❌ WRONG: ["( Turck-Chieze 2005 )"]

    Snippet: "Smith et al. ( 2003 ) noted a correspondence between the cessation of high latitude CMEs and the polar field reversal."
    Substring: ["Smith et al. ( 2003 )"]

    Snippet: "In work by Jones (2000a,b) it was demonstrated that dark matter detection can be elegantly detected."
    Substring: ["Jones (2000a,b)"]

    Snippet: This is a process that should conclude within millions of years (e.g., Guillot et al. 1996 ).
    Substring: ["(e.g., Guillot et al. 1996 )"]

    Snippet: "This paper describes the cluster (and assumes similar principles; see Boboni et al. ( 2011 ) where this assumption is relaxed)."
    Substring: ["Boboni et al. ( 2011 )"]

    Snippet: "This is consistent with previous results (Pereira et al., in preparation)."
    Substring: ["(Pereira et al., in preparation)"]

    Snippet: "More imaging showed significant expansion (Jones, Ryan, et al., submitted)"
    Substring: ["(Jones, Ryan, et al., submitted)"]

    Snippet: "It also hosts a Compton-thick AGN in the Western component ( Smith 2004 ), observed directly in hard X-rays (Della Ceca et al. 2002 ; Ballo et al. 2004 ).
    Substring: ["( Smith 2004 )", "(Della Ceca et al. 2002 ; Ballo et al. 2004 )"]

    Snippet: "Even prior to the Ballo et al. (2013 ) result on the potential rotationally supported plane in M31..."
    Substring: ["Ballo et al. (2013 )"]

    Snippet: "Figure 3. Effects on N-body mass functions of different group-finding schemes."
    Substring: []

    ##
    Keeping the rules and examples above in mind, identify the citation substrings in the following snippet:

  \end{Verbatim}
\end{promptbox}

\begin{promptbox}[Create Citation Tuples Prompt]
  \begin{Verbatim}[breaklines=true, fontsize=\small]
    Task:
    Given a single input string containing one or more inline research citations, extract and output them as a JSON array of [author(s), year] pairs. Follow these rules exactly:

    1. Match patterns where an author block is followed by a 4-digit year, optionally suffixed by letters (e.g. 2008a, 2008b).
    2. Author blocks may include letters, spaces, commas, ampersands (&), "et al.", hyphens, and accented characters.
    3. Only split into separate citations at:
    - Semicolons (";")
    - Commas or spaces immediately preceding a 4-digit year (with optional letter suffix)
    - Parentheses or brackets around the citation—but ignore parentheses after matching.
    4. Do NOT split author names joined by "and" or "&". Do NOT split consecutive author names (e.g. Smith Jones 1997, van Hoek Smith 1997)
    5. For entries with multiple years after the same author block (e.g. "Smith 1997, 1998"), output one citation per year, reusing the same author block.
    6. Discard any citations without a 4-digit year, or those indicating "submitted", "in preparation", etc.
    7. Output ONLY the JSON array (no extra text or explanation).

    Examples:
    """
    Input: "Ussiri and Lal, 2013"
    Output: [["Ussiri and Lal", "2013"]]

    Input: "Ussiri Lal 2013"
    Ouptut: [["Ussiri Lal", "2013"]]

    Input: "Caffau et al. (2008a)"
    Output: [["Caffau et al.", "2008"]]

    Input: "Sweatman & Young, 2000"
    Output: [["Sweatman & Young", "2000"]]

    Input: "Sweatman and Young, 2000"
    Output: [["Sweatman and Young", "2000"]]

    Input: "van Der Monde 1997"
    Output: [["van Der Monde", "1997"]]

    Input: "Turck-Chieze(2005)"
    Output: [["Turck-Chieze", "2005"]]

    Input: "Jones (2000a,b)"
    Output: [["Jones", "2000a"], ["Jones", "2000b"]]

    Input: "Caffau and Brun (2009)"
    Output: [["Caffau and Brun", "2009"]]

    Input: "Smith and Jones (2003)"
    Output: [["Smith and Jones", "2003"]]

    Input: "Smith & Jones (2003)"
    Output: [["Smith & Jones", "2003"]]

    Input: "Smith Jones (2003)"
    Output: [["Smith Jones", "2003"]]

    Input: "Brun, Turck-Chièze Zahn 1999 , Charbonnel Talon 2005"
    Output: [["Brun, Turck-Chièze Zahn", "1999"], ["Charbonnel Talon", "2005"]]

    Input: "Brun, Turck-Chièze Zahn 1999 ; Charbonnel Talon 2005"
    Output: [["Brun, Turck-Chièze Zahn", "1999"], ["Charbonnel Talon", "2005"]]

    Input: "Fabian 1999 ; Di Matteo, Wilman Crawford 2002 ; King 2003 , 2005"
    Output: [["Fabian", "1999"], ["Di Matteo, Wilman Crawford", "2002"], ["King", "2003"], ["King", "2005"]]

    Input: "Smith et al. (2002)", "Jones (2004)"
    Output: [["Smith et al.", "2002"], ["Jones", "2004"]]

    Input: "Audouse Silk 1995 ; Ryan, Norris Beers 1996"
    Output: [["Audouse Silk", "1995"], ["Ryan, Norris Beers", "1996"]]

    Input: "This is consistent with previous results (Pereira et al., in preparation)."
    Output: []

    Input: "More imaging showed significant expansion (Jones, Ryan, et al., submitted)"
    Output: []
    """

    Extract the inline citations from the following text:
    Input:

  \end{Verbatim}
\end{promptbox}

\subsection{Document Representation by `Contributions'} \label{app:prompts}

\begin{promptbox}[Contribution Extraction Prompt - v1]
  \begin{Verbatim}[breaklines=true, fontsize=\small]
    You are an expert research assistant. Summarize the original scientific findings of the following research paper. Then write them out as a list of strings.

    ## Task
    - Write the findings as a JSON array of strings, each finding being one string element.
    - Do not include acknowledgments or references: focus only on the original research contributions made in this paper
    - Only write out the JSON array, do not include any other text or formatting.

    Example output:
    {{
    findings: ["Contribution 1", "Contribution 2", "Contribution 3"]
    }}

    ## Paper:
    {paper}
  \end{Verbatim}
\end{promptbox}

\begin{promptbox}[Contribution Extraction Prompt - v2]
  \begin{Verbatim}[breaklines=true, fontsize=\small]
    You are given the full text of a single scientific research paper. Extract all factual statements asserted or utilized in this paper, including both novel contributions and descriptions of methods/instruments/data used.

    Hard requirements:

    1. Output ONLY a single JSON object: {{"findings":[ ... ]}}. No extra text, no commentary, no additional keys.
    2. Each array item must be a single, self-contained declarative sentence (one idea per sentence). Avoid pronouns and author-phrases (no "we", "authors", "found", "concluded", "determined", "observed"). State the fact directly (e.g., "Stellar bulge contains significant population of young stars").
    3. Ensure each statement is fully interpretable on its own. Include definitions or clarifying context for all domain-specific terms, abbreviations, or categories:
    - Include the subject and scope (e.g., "in the galactic Bulge", "in sample of 200 patients", "in dataset X", "in cohort aged 50–70").
    - If the sentence refers to a term defined within the paper, briefly include the definition or context so it is fully interpretable out of context
    - If the sentence refers to a method/analysis, name the method and general dataset characteristics (e.g., "High-resolution spectroscopy of F/G dwarf stars").
    - Focus on general methodological approaches, broad findings, and conceptual contributions rather than specific numeric values. Use descriptive terms like "significant", "substantial", "moderate", "enhanced", "reduced" instead of precise measurements.
    3. Ensure each statement is fully interpretable on its own. Include definitions or clarifying context for all domain-specific terms, abbreviations, or categories:
    - Include the subject and scope (e.g., "in the galactic Bulge", "in sample of 200 patients", "in dataset X", "in cohort aged 50–70").
    - If discussing a subset, subgroup, or category, ALWAYS include what it's a subset OF.
    - BAD: "Outliers have masses from 9.4 to 11.1"
    - GOOD: "Galaxies offset from mass-metallicity relation (MZR) have masses from 9.4 to 11.1"
    - If the sentence refers to a term defined within the paper, briefly include the definition or context so it is fully interpretable out of context.
    - For well-established acronyms and abbreviations in the field, include BOTH the full term and acronym on first use:
    - Use the format: "full term (ACRONYM)"
    - Examples: "mass-metallicity relation (MZR)", "cosmic microwave background (CMB)", "supermassive black hole (SMBH)", "active galactic nucleus (AGN)", "asymptotic giant branch (AGB)", "second generation (2G)"
    - Apply this to: major concepts, standard measurements, established relations, common astrophysical objects, and widely-recognized instruments
    - If a term appears in multiple statements, include the acronym definition in each statement (since each must be independently interpretable)
    - If the sentence refers to a method/analysis, name the method and general dataset characteristics (e.g., "High-resolution spectroscopy of F/G dwarf stars").
    - Focus on general methodological approaches, broad findings, and conceptual contributions rather than specific numeric values. Use descriptive terms like "significant", "substantial", "moderate", "enhanced", "reduced" instead of precise measurements.
    4. Prefer concise, general statements ($\le 160$ characters when possible) that capture the essence of findings.
    5. Extract these types of statements (when present):
    - original findings/contributions
    - methodologies & experimental design (instruments, general datasets, analytical approaches)
    - qualitative and broad quantitative results
    - instruments/tools used (even if established), with their purpose and capabilities
    - conclusions, interpretations, and statements motivated by the paper's findigns
    - method-observation pairs (e.g. [Method X] shows/detects/etc. [Observation Y])
    - limitations/caveats/assumptions
    - new resources (data/code/models)
    - noteworthy comparisons to prior work
    - open questions, unresolved problems, or acknowledged uncertainties. Prefix them with "Open question:"
    6. For each statement, write out findings at MULTIPLE levels of specificity:
    - High-level interpretations: a broader, more general, citation-style statement
    - Mid-level: General findings with key qualifiers ($\le 160$  chars)
    - Detailed: Specific conditions, parameters, and quantitative results
    - If a finding can have multiple high-level interpretations, write out all of them
    - ALWAYS write statements with complete context
    7. Before finalizing each statement, verify: "If I gave this sentence to someone unfamiliar with the paper, would they understand what population/sample/reference frame is being discussed?" If no, add the missing context.

    ## Research Paper

      {paper}

    ## Example expected output:

    {{
    "findings": ["finding 1...", "finding 2...", ...]
    }}

  \end{Verbatim}
\end{promptbox}

\end{document}